# A Statistical Peek into Average Case Complexity

NIRAJ KUMAR SINGH, Birla Institute of Technology Mesra
SOUBHIK CHAKRABORTY, Birla Institute of Technology Mesra
DHEERESH KUMAR MALLICK, Birla Institute of Technology Mesra

The present paper gives a statistical adventure towards exploring the average case complexity behavior of computer algorithms. Rather than following the traditional count based analytical (pen and paper) approach, we instead talk in terms of the weight based analysis that permits mixing of distinct operations into a conceptual bound called the statistical bound and its empirical estimate, the so called "empirical O". Based on careful analysis of the results obtained, we have introduced two new conjectures in the domain of algorithmic analysis. The analytical way of average case analysis falls flat when it comes to a data model for which the expectation does not exist (e.g. Cauchy distribution for continuous input data and certain discrete distribution inputs as those studied in the paper). The empirical side of our approach, with a thrust in computer experiments and applied statistics in its paradigm, lends a helping hand by complimenting and supplementing its theoretical counterpart. Computer science is or at least has aspects of an experimental science as well, and hence hopefully, our statistical findings will be equally recognized among theoretical scientists as well.

## 1. INTRODUCTION

Traditionally the analysis of an algorithm's efficiency is done through its mathematical analysis. Although these techniques can be applied successfully to many simple algorithms, the power of mathematics, even when enhanced with more advanced techniques, is far from limitless [1][2][3]. Robustness of the analytical approach itself can be challenged on the ground that even some seemingly simple algorithms have proved to be very difficult to analyze with mathematical precision and certainty [4]. This is especially true for average case analysis.

Average case complexity analysis is an important idea as it explains how certain algorithms with bad worst case complexity perform better on the average. The principal alternative to the conventional mathematical analysis of an algorithm's efficiency is its empirical analysis [4]. Recently there has been an upswing in interest in experimental work in theoretical computer science community because of growing recognition that theoretical results cannot tell the full story about real-world algorithmic performance [5]. Empirical researchers have serious objection to applying mathematical bounds in average case complexity analysis [6] [7] [8]. Indeed, it was the lack of scientific rigor in early experimental work that led Knuth and other researchers in the 1960's to emphasize worst- and- average case analysis and the more general conclusions they could provide, especially with respect to asymptotic behavior [5]. Empirical analysis is capable of finding patterns inside a pattern exhibited by its theoretical counterpart. Its result in turn may reinvigorate the theoretical establishments by complimenting and supplementing the already known theoretical findings.

Here, through this paper we suggest for an alternative tool, the so called 'statistical bound' and its empirical estimate (empirical-O) [6] [7]. The performance guarantee is perhaps the biggest strength of the mathematical worst case bound. Such a guarantee can even be ensured with empirical-O (denoted as $O_{emp}$) by verifying the complexity robustness across the commonly used and standard data distributions inputs. The statistical approach is well equipped with tools and techniques which, if used scientifically, can provide a reliable and valid measure for an algorithm's complexity. See [9] [10] for interesting discussions on the statistical analysis of experiments in a rigorous fashion. For more on statistical bounds and empirical-O see the Appendix.

Average case inputs typically correspond to randomly obtained samples. Such random sequences may constitute certain well known data patterns or sometimes may even result in unconventional models. We in practice are not very much concerned with best case analysis as it often paints a very optimistic picture. In this paper we confine ourselves to average case analysis only as we find it practically more exciting than the others.

Our statistical adventure explores the average case behavior of the well known standard quick sort algorithm [11] as a case study. We find it a suitable one as it exhibits a significant performance gap between its theoretical average and worst case behaviors. Our analysis introduces a phrase "pseudo linear complexity" which we obtain against the theoretical $O(n\log_2 n)$ complexity for our candidate algorithm. We also talk about the rejection of the average case robustness claim for quick sort algorithm made by the theoretical scientists.

## 2. AVERAGE CASE ANALYSIS USING STATISTICAL BOUND ESTIMATE OR 'EMPIRICAL-O'

The average case analysis was done by directly working on program run time to estimate the weight based statistical bound over a finite range by running computer experiments [12] [13]. This estimate is called empirical-O. Here time of an operation is taken as its weight. Weighing allows collective consideration of all operations, trivial or non-trivial, into a conceptual bound. We call such a bound a statistical bound opposed to the traditional count based mathematical bounds which is operation specific. Since the estimate is obtained by supplying numerical values to the weights obtained by running computer experiments, the credibility of this bound estimate depends on the design and analysis of computer experiment in which time is the response. The interested reader is suggested to see [6] [14] to get more insight into statistical bounds and empirical-O. Average complexity is traditionally (by a count based analysis using pen and paper) found by applying mathematical expectation to the dominant operation. If the dominant operation is wrongly chosen (this is likely in a complex code or even in a simple code in which a dominant operation comes fewer times as compared to a less dominant operation which comes more; as for example, in Amir Schoor's n-by-n matrix multiplication algorithm $n^2$ comparisons dominate over multiplications indicating an empirical $O(n^2)$ complexity as the statistical bound estimate while Schoor applied mathematical expectation on multiplication and obtained quite a different result, namely, an average $O(d_1 d_2 n^3)$ complexity, where $d_1$ and $d_2$ are the fraction of non-zero elements of the pre and post factor matrices [6][7]. Second, the robustness can also be challenged as the probability distribution over which expectation is taken may not be realistic over the problem domain.

This section includes the empirical results obtained for average case analysis of quick sort algorithm. The samples are generated randomly, using random number generating function, to characterize discrete uniform distribution models with K as its parameter. Our sample sizes lie in between $5*10^5$ and $5*10^6$. The discrete uniform distribution depends on the parameter K [1 .......K], which is the key to decide the range of sample obtained.

Most of the mean time entries (in seconds) are averaged over 500 trial readings. These trial counts however, should be varied depending on the extent of "noise" present at a particular 'n' value. As a rule of thumb, greater the "noise" at each point of 'n', more should be the numbers of observations.

***System specification:*** All the computer experiments were carried out using PENTIUM 1600 MHz processor and 512 MB RAM. Statistical models/results are obtained using Minitab-15 statistical package. The standard quick sort is implemented using C language by the authors themselves.

It should be understood that although program run time is system dependent, we are interested in identifying *patterns* in the run time rather than run time itself. It may be emphasized here that statistics is the science of identifying and studying *patterns* in numerical data related to some problem under study.

**2.1 Average case analysis over discrete uniform inputs (case study-1)**

In our first case study we observed the mean times for specific sized input data over the entire range for different K values. The observed data is recorded in table 1.

Table 1. Observed mean times in second(s)

| n↓ K→ | 50 | 500 | 5000 | 10000 | 20000 | 25000 | 50000 | 500000 | 5000000 |
|---|---|---|---|---|---|---|---|---|---|
| 500000 | 12.688 | 1.407 | 0.33104 | 0.26416 | 0.24652 | 0.23976 | 0.24512 | 0.24276 | 0.22824 |
| 1000000 | 50.703 | 5.219 | 0.80956 | 0.57828 | 0.47392 | 0.45504 | 0.42136 | 0.42512 | 0.42568 |
| 1500000 | 114.156 | 11.531 | 1.56372 | 1.02256 | 0.78744 | 0.75248 | 0.66752 | 0.66304 | 0.66576 |
| 2000000 | 203.546 | 20.359 | 2.57132 | 1.61632 | 1.18136 | 1.11132 | 0.96184 | 0.97436 | 0.96308 |
| 2500000 | 317.562 | 31.657 | 3.8498 | 2.33948 | 1.65008 | 1.5251 | 1.3012 | 1.30492 | 1.29804 |
| 3000000 | 457.797 | 45.437 | 5.3688 | 3.1814 | 2.17812 | 2.0016 | 1.6703 | 1.68316 | 1.66928 |
| 3500000 | *** | 61.719 | 7.156 | 4.1607 | 2.76612 | 2.5157 | 2.0859 | 2.09012 | 2.08508 |
| 4000000 | *** | 80.484 | 9.187 | 5.2672 | 3.4406 | 3.1045 | 2.5217 | 2.53112 | 2.53236 |
| 4500000 | *** | 101.875 | 11.43233 | 6.5229 | 4.17188 | 3.7674 | 3.0236 | 3.0282 | 3.01748 |
| 5000000 | *** | 126.078 | 13.93267 | 7.890333 | 4.9515 | 4.4328 | 3.5342 | 3.5453 | 3.54188 |

A careful look at table 1 reveals that for smaller 'K' values we get quadratic complexity models. Our point is further strengthened when we go through the statistical data of tables 2(A-H). So we can safely write $Y_{avg}(n)=O_{emp}(n^2)$, at least for K≤10000. It must be kept in mind that the associated constant term in this inequality is not a generic one, rather a constant dependent on the range of input size. However, if the system invariance property of $O_{emp}$ is ensured then this value may be treated as a constant across the systems provided the range of input size is kept fixed.

Table 2(A) Regression Analysis: y versus n, nlgn, n^2 for [k=500]

```
The regression equation is
y = - 0.541 + 0.000013 n - 0.000001 nlgn + 0.000000 n^2

Predictor        Coef       SE Coef       T      P
Constant       -0.5407       0.3653    -1.48   0.189
n            0.00001267   0.00000570    2.22   0.068
nlgn        -0.00000061   0.00000027   -2.23   0.067
n^2          0.00000000   0.00000000   62.14   0.000

S = 0.0876502   R-Sq = 100.0%   R-Sq(adj) = 100.0%

Analysis of Variance

Source          DF       SS       MS          F        P
Regression       3   16607.5   5535.8   720573.28   0.000
Residual Error   6       0.0      0.0
Total            9   16607.6

Source  DF   Seq SS
n        1   15770.1
nlgn     1     807.8
n^2      1      29.7
```

Table 2(B). Regression Analysis: y versus n, nlgn for [k=500]

**Regression Analysis: y versus n, nlgn**

```
The regression equation is
y = 19.9 - 0.000335 n + 0.000016 nlgn

Predictor        Coef       SE Coef       T      P
Constant       19.927         3.712     5.37   0.001
n          -0.00033492   0.00002628   -12.74   0.000
nlgn        0.00001598   0.00000116    13.80   0.000

S = 2.06007   R-Sq = 99.8%   R-Sq(adj) = 99.8%

Analysis of Variance

Source          DF       SS       MS         F       P
Regression       2   16577.9   8288.9   1953.16   0.000
Residual Error   7      29.7      4.2
Total            9   16607.6

Source  DF   Seq SS
n        1   15770.1
nlgn     1     807.8
```

Table 2(C). Regression Analysis: y versus n, nlgn, n^2 for [k=5000]

**Regression Analysis: y versus n, nlgn, n^2 for [k=5000]**

The regression equation is
y = 0.242 - 0.000003 n + 0.000000 nlgn + 0.000000 n^2

Predictor        Coef         SE Coef      T       P
Constant         0.24156      0.03749      6.44    0.001
n               -0.00000281   0.00000059  -4.80    0.003
nlgn             0.00000015   0.00000003   5.24    0.002
n^2              0.00000000   0.00000000  53.27    0.000

S = 0.00899539   R-Sq = 100.0%   R-Sq(adj) = 100.0%

Analysis of Variance

Source          DF    SS         MS        F            P
Regression       3   198.024    66.008    815749.05    0.000
Residual Error   6     0.000     0.000
Total            9   198.024

Source    DF    Seq SS
n          1    189.642
nlgn       1      8.153
n^2        1      0.230

Table 2(D). Regression Analysis: y versus n, nlgn, n^2 for [k=10000]

The regression equation is

y = 0.0982 - 0.000000 n + 0.000000 nlgn + 0.000000 n^2

Predictor        Coef         SE Coef      T       P
Constant         0.09823      0.02367      4.15    0.006
n               -0.00000019   0.00000037  -0.51    0.625
nlgn             0.00000002   0.00000002   1.18    0.284
n^2              0.00000000   0.00000000  47.33    0.000

S = 0.00567979   R-Sq = 100.0%   R-Sq(adj) = 100.0%

Analysis of Variance

Source          DF    SS        MS        F            P
Regression       3   61.649    20.550    636996.80    0.000
Residual Error   6    0.000     0.000
Total            9   61.649

Source    DF    Seq SS
n          1    59.348
nlgn       1     2.228
n^2        1     0.072

Table 2(E). Regression Analysis: y versus n, nlgn, n^2 for [k=20000]

**Regression Analysis: y versus n, nlgn, n^2 for [k=20000]**

The regression equation is

```
y = 0.160 - 0.000002 n + 0.000000 nlgn + 0.000000 n^2

Predictor        Coef      SE Coef       T       P
Constant      0.16047      0.02249    7.13   0.000
n          -0.00000150   0.00000035   -4.27  0.005
nlgn        0.00000009   0.00000002    5.11  0.002
n^2         0.00000000   0.00000000   21.61  0.000

S = 0.00539657   R-Sq = 100.0%   R-Sq(adj) = 100.0%

Analysis of Variance

Source          DF      SS        MS         F         P
Regression       3   23.4474   7.8158   268372.19   0.000
Residual Error   6    0.0002   0.0000
Total            9   23.4476

Source  DF    Seq SS
n        1    22.8198
nlgn     1     0.6140
n^2      1     0.0136
```

Table 2(F). Regression Analysis: y versus n, nlgn, n^2 for [k=25000]

```
The regression equation is

y = 0.129 - 0.000001 n + 0.000000 nlgn + 0.000000 n^2

Predictor        Coef      SE Coef       T       P
Constant      0.12939      0.04199    3.08   0.022
n          -0.00000103   0.00000066   -1.57  0.168
nlgn        0.00000006   0.00000003    2.03  0.089
n^2         0.00000000   0.00000000    9.98  0.000

S = 0.0100740   R-Sq = 100.0%   R-Sq(adj) = 100.0%

Analysis of Variance

Source          DF      SS        MS         F        P
Regression       3   18.5835   6.1945   61038.16   0.000
Residual Error   6    0.0006   0.0001
Total            9   18.5841

Source  DF    Seq SS
n        1    18.1414
nlgn     1     0.4320
n^2      1     0.0101
```

Table 2(G). Regression Analysis: y versus n, nlgn, n^2 for [k=50000]

```
The regression equation is
```

```
y = 0.178 - 0.000002 n + 0.000000 nlgn + 0.000000 n^2

Predictor         Coef      SE Coef       T       P
Constant       0.17829      0.02566    6.95   0.000
n          -0.00000161   0.00000040   -4.03   0.007
nlgn        0.00000009   0.00000002    4.75   0.003
n^2         0.00000000   0.00000000    9.05   0.000

S = 0.00615659   R-Sq = 100.0%   R-Sq(adj) = 100.0%

Analysis of Variance

Source           DF       SS        MS           F         P
Regression        3   11.4303    3.8101   100521.12    0.000
Residual Error    6    0.0002    0.0000
Total             9   11.4306

Source   DF    Seq SS
n         1   11.2128
nlgn      1    0.2144
n^2       1    0.0031
```

Table 2(H). Regression Analysis: y versus n, nlgn for [k=50000]

```
The regression equation is

y = 0.388 - 0.000005 n + 0.000000 nlgn

Predictor         Coef      SE Coef       T       P
Constant       0.38768      0.03931    9.86   0.000
n          -0.00000517   0.00000028  -18.57   0.000
nlgn        0.00000026   0.00000001   21.22   0.000

S = 0.0218164   R-Sq = 100.0%   R-Sq(adj) = 100.0%

Analysis of Variance

Source           DF       SS        MS          F         P
Regression        2   11.4272    5.7136   12004.53    0.000
Residual Error    7    0.0033    0.0005
Total             9   11.4306

Source   DF    Seq SS
n         1   11.2128
nlgn      1    0.2144
```

"Even if you find a low $r^2$ value in an analysis, make sure to go back and look at the regression coefficients and their t values. You may find that, despite the low $r^2$ value, one or more of the regression coefficients is still strong and relatively well known. In the same manner, a high $r^2$ value doesn't necessarily mean that the model that you have fitted to the data is the right model. That is, even when $r^2$ is very large, the fitted model may not accurately predict the response. It's the job of lack of fit or goodness of fit tests to decide if a model is a good fit to the data [15]".

*Justification for preferring quadratic model over $nlog_2n$ for K=500:*
Statistical data in tables 2(A, B) justify the choice of quadratic model over $nlog_2n$. The standard error is reduced from 2.06007 to 0.0876502 when we go for quadratic model. Although a very slight improvement in $r^2$ data is observed, the F value (720573.28) for quadratic model is much higher than the corresponding value (1953.16) for $nlog_2n$ curve.

*Justification for preferring $nlog_2n$ model over quadratic one for K=50000:*
As we move from smaller to higher k values there is a significant gradual decrement in t statistic of $n^2$ term which is obvious from statistics given in tables 2(A-G). Ultimately for the specified K value the t statistic of $nlog_2n$ approaches close to that of $n^2$ term in the model.

For a given range of input size (in our case it is $5*10^5$ to $5*10^6$) an increment in K beyond a threshold (indeed a range) ensures its best performance for random inputs: i.e. $Y_{avg}(n) = O(nlog_2n)$. So it is the K value of sample which decides on the average case behavior of algorithms (quick sort in particular). This information is important as its prior knowledge may influence the choice for a particular algorithm in advance.

Our study refutes the robustness claim made for $nlog_2n$ average case behavior of quick sort. See reference [16] for an interesting discussion on the robustness of average complexity measure.

**2.2 Average case analysis over discrete uniform inputs (case study-2)**

The frequency of occurrence of a particular element $e_i$ belonging to a sample is its tie density $t_d(e_i)$. As we are dealing with uniform distribution samples only, to enhance the readability, we simply drop the bracketed term and hence $t_d$ corresponds to the tie density of each element in the sample. For an interesting discussion on the effect of tied elements on the algorithmic performance the reader is referred to [17].

### 2.2.1 Statistical results and its analysis

In this section our study is focused around analyzing algorithmic performance when it is subjected to uniform distribution data coupled with constant probability of tied elements. The observed mean time is recorded in table 3. The corresponding statistical analysis result is presented in tables 4(A-B). This result clearly suggests an $O_{emp}(n)$ complexity across the various tie density values. The resulting residual plots for response are given in figures 1&2.

Table 3. Observed mean times in seconds

| n↓ $t_d$→ | 1 | 10 | 100 | 1000 | 10000 | 100000 |
|---|---|---|---|---|---|---|
| 500000 | 0.24276 | 0.24512 | 0.33104 | 1.407 | 12.688 | *** |
| 1000000 | 0.4316 | 0.4283 | 0.5734 | 2.7594 | 25.219 | *** |
| 1500000 | 0.6732 | 0.664 | 0.8561 | 4.1281 | 37.828 | *** |
| 2000000 | 0.964 | 0.9672 | 1.1828 | 5.5125 | 50.375 | *** |
| 2500000 | 1.2999 | 1.2952 | 1.5062 | 6.9016 | 63 | *** |
| 3000000 | 1.6767 | 1.6703 | 1.7781 | 8.2842 | 76.016 | *** |
| 3500000 | 2.0829 | 2.0859 | 2.0844 | 9.7 | 88.219 | *** |
| 4000000 | 2.5391 | 2.5343 | 2.5391 | 11.10433 | 100.828 | *** |
| 4500000 | 3.0219 | 3.0186 | 3.0187 | 12.51567 | 113.468 | *** |
| 5000000 | 3.54188 | 3.5453 | 3.5342 | 13.93267 | 126.078 | *** |

Table 4(A). Regression Analysis: t versus n, $n\log_2 n$ for $t_d=1$

```
The regression equation is
t = 0.390 - 0.000005 n + 0.000000 nlogn

Predictor         Coef       SE Coef        T       P
Constant       0.38954       0.04154     9.38   0.000
n          -0.00000517   0.00000029   -17.57   0.000
nlogn       0.00000026   0.00000001    20.08   0.000

S = 0.0230505   R-Sq = 100.0%   R-Sq(adj) = 100.0%

PRESS = 0.0164288   R-Sq(pred) = 99.86%

Analysis of Variance

Source            DF        SS        MS          F       P
Regression         2   11.4483    5.7241   10773.36   0.000
Residual Error     7    0.0037    0.0005
Total              9   11.4520

Source  DF    Seq SS
n        1   11.2341
nlogn    1    0.2142

Obs    n            t       Fit    SE Fit   Residual   St Resid
 1   500000  0.24276   0.26984   0.01955   -0.02708   -2.22R
 2  1000000  0.43160   0.41036   0.01181    0.02124    1.07
 3  1500000  0.67320   0.64910   0.01037    0.02410    1.17
 4  2000000  0.96400   0.95163   0.01085    0.01237    0.61
 5  2500000  1.29990   1.30158   0.01092   -0.00168   -0.08
 6  3000000  1.67670   1.68933   0.01025   -0.01263   -0.61
 7  3500000  2.08290   2.10852   0.00942   -0.02562   -1.22
 8  4000000  2.53910   2.55461   0.00969   -0.01551   -0.74
 9  4500000  3.02190   3.02423   0.01225   -0.00233   -0.12
10  5000000  3.54188   3.51474   0.01702    0.02714    1.75

R denotes an observation with a large standardized residual.
```

Table 4(B). Regression Analysis: t versus n, $n\log_2 n$ for $t_d=1000$

```
Regression Analysis: t versus n, nlogn [td=1000]

The regression equation is
```

```
t = 0.106 + 0.000002 n + 0.000000 nlogn

Predictor        Coef      SE Coef        T       P
Constant     0.105638    0.008300    12.73   0.000
n           0.00000169  0.00000006    28.82   0.000
nlogn       0.00000005  0.00000000    18.59   0.000

S = 0.00460600   R-Sq = 100.0%   R-Sq(adj) = 100.0%

PRESS = 0.000254515   R-Sq(pred) = 100.00%

Analysis of Variance

Source          DF       SS        MS          F        P
Regression       2   160.103    80.051   3773304.32  0.000
Residual Error   7     0.000     0.000
Total            9   160.103

Source   DF    Seq SS
n         1   160.096
nlogn     1     0.007

Obs    n           t       Fit     SE Fit   Residual   St Resid
 1   500000    1.4070   1.4082   0.0039    -0.0012     -0.51
 2  1000000   2.7594   2.7590   0.0024     0.0004      0.10
 3  1500000   4.1281   4.1279   0.0021     0.0002      0.04
 4  2000000   5.5125   5.5087   0.0022     0.0038      0.94
 5  2500000   6.9016   6.8982   0.0022     0.0034      0.84
 6  3000000   8.2842   8.2947   0.0020    -0.0105     -2.54R
 7  3500000   9.7000   9.6970   0.0019     0.0030      0.71
 8  4000000  11.1043  11.1043   0.0019     0.0000      0.01
 9  4500000  12.5157  12.5160   0.0024    -0.0003     -0.07
10  5000000  13.9327  13.9315   0.0034     0.0012      0.38

R denotes an observation with a large standardized residual.
```

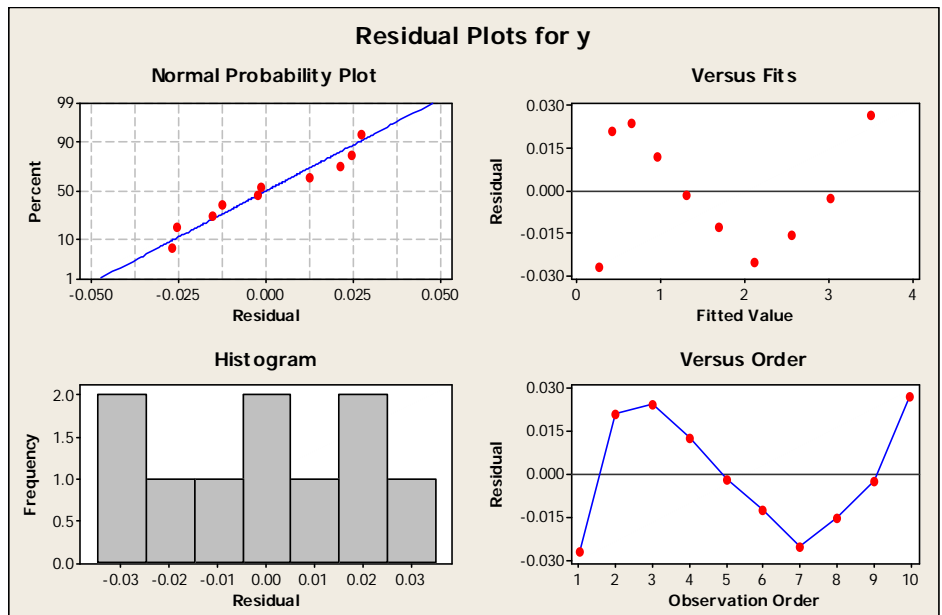

Fig.1. Residual plot corresponding to $t_d=1$

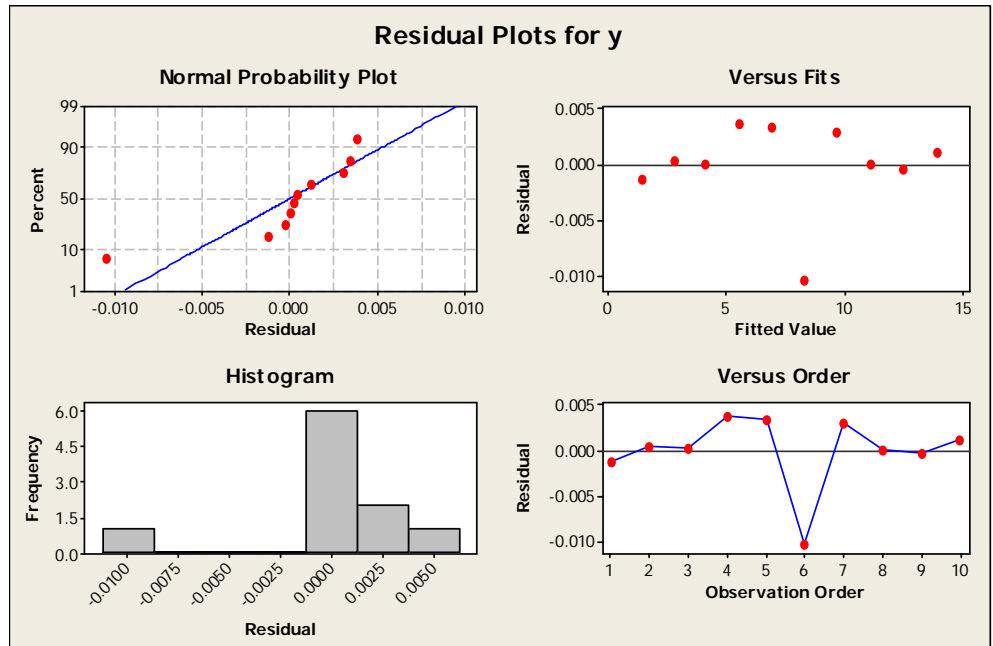
Fig.2. Residual plot corresponding to $t_d=1000$

With a value of 28.82 the t statistic is significantly higher for linear term against the value of 18.59 for $nlog_2n$ term in the later model. Opposed to this the t value for $nlog_2n$ term (20.08) is higher than that of linear term (-17.57) in the earlier model. This observation led us to conclude that beyond a threshold value of tie density we expect linear patterns rather than $nlog_2n$.

***Pseudo linear complexity model:*** Analyzing algorithmic complexity through the study of growth patterns is an important idea, but things could be different when it comes to practice. Unlike theoretical analysis an empirical analysis is essentially done over a feasible finite range. Hence, while going for empirical analysis, one should not always rely completely on the growth pattern as even the individual time values can have their own share (sometimes even major) to contribute in deciding the final complexity of the program in question. A careful observation of table 3 would further clear this point. The CPU times are more or less comparable when we look into the columns (table 3) for $t_d$=1, 10, and 100. However, as we moved further for higher $t_d$ values, the timing differences with respect to the CPU times measured against the unit $t_d$ value became prominent. The tables 1 & 3 are related by the fact that a column in table 3 would correspond to a rightward diagonal in Table 1 (if all the relevant entries were present).

Each point on an average complexity curve, obtained for some higher $t_d$ (say 1000 as in fig. 3) value, gives an upper bound to the corresponding point (sample size) present in some quadratic curve (in fig. 3 these curves correspond to k=5000 and 10000) obtained as complexity model for essentially similar input categories. Here we must remember that the tie density cannot be an arbitrary number as it can always be limited by the sample size N. Although, the actual timings are compared among the models obtained from possibly different data patterns they all belong to the very same family of inputs (average case inputs). Hence, although following linear patterns we call such a model as a "*pseudo linear complexity model*".

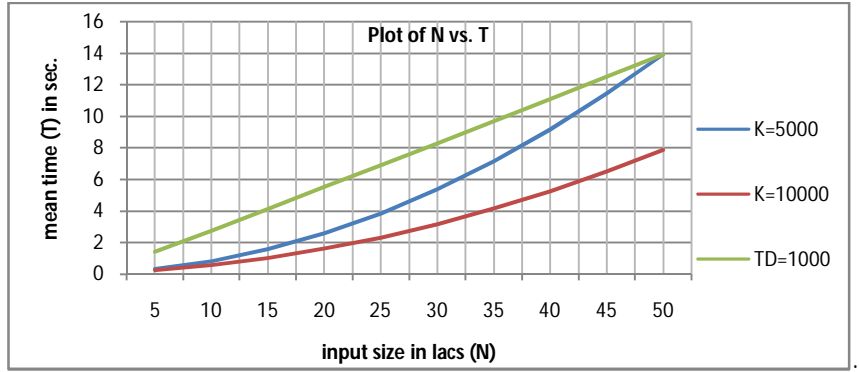

Fig. 3. Relative curves demonstrating pseudo linear complexity model

At this point we are in a position to claim that: "*The uniform distribution data with similar (at least theoretically) density of tied elements is guaranteed to yield $O_{emp}(n)$ growth rate*". In context of theoretical analysis this result is quite unexpected as even the best case theoretical count happens to be $\Omega(n\log_2 n)$ and not a linear one. Based on our analysis result we put our points in the form of the following two conjectures.

*Conjecture 1:* Over uniform distribution data with similar density of tied elements, a theoretical $O(n\log_2 n)$ complexity approaches towards an empirical $O_{emp}(n)$ complexity for average case inputs having sufficiently large $t_d$ values.

*Conjecture 2:* As the sample tie density $t_d$ goes beyond a certain threshold value $t_{dt}$, (i.e. for all $t_d > t_{dt}$ ) even the seemingly linear complexity model, as claimed in conjecture 1, is found to be quadratic in practice, provided the sample range remains the same. And hence, we call such a linear model as "a pseudo linear complexity model".

Although, the presence of other reasons cannot be ruled out, our failure in identifying the dominant operation(s), and that too correctly, present inside the code is among the reasons for the observed behavior.

**Theoretical justification:** It is well known that runtime of quick sort depends on the number of distinct elements [18], which in this paper, is reflected in the parameter 'k'. If $t_d$ is the tie density, then $n=k*t_d$. With fixed $t_d$, in a feasible finite range setup the value of 'k' will increase linearly with n. The similar argument is equally applicable for fixed 'k' value (see fig. 4 A & B). Also for fixed sample sizes the response is maximum when $t_d=n$ ($t_d$ cannot exceed n), which is the case when all elements are same valued. The response is minimum when $t_d=1$, (i.e. n=k) when all elements are distinct (at least theoretically).

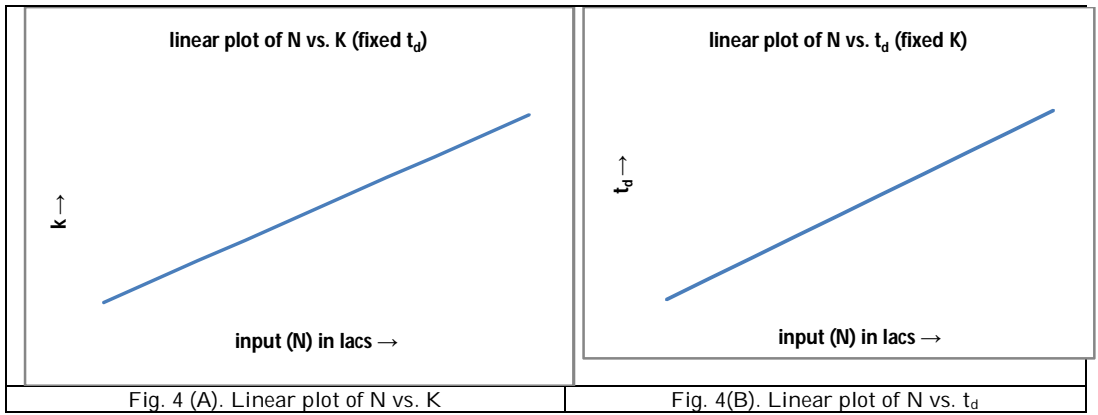

| Fig. 4 (A). Linear plot of N vs. K | Fig. 4(B). Linear plot of N vs. $t_d$ |

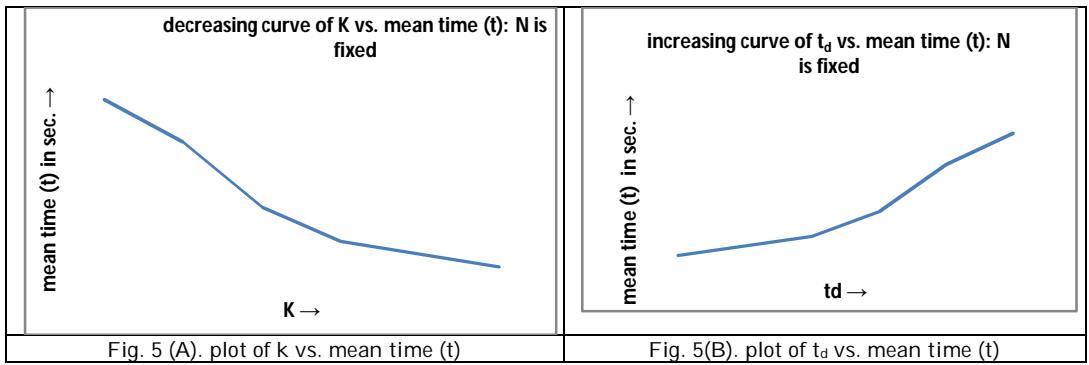

| Fig. 5 (A). plot of k vs. mean time (t) | Fig. 5(B). plot of $t_d$ vs. mean time (t) |

The expected behavior of a random sample is $O(n\log_2 n)$ complexity whereas, for a fixed sample size the run time of quick sort is a decreasing function over parameter 'k' (see fig. 5 A & B). This observation affects the runtime of the samples decreasing the overall runtime from $n\log_2 n$ complexity to a linear one.

Following this discussion it seems that the empirical analysis has the potential to cross the barrier, which otherwise is not attainable through its theoretical counterpart.

*2.2.2. Quick sort for unusual data model:*

Analyzing the algorithmic behavior, for average case performance, through analytical approach has its own inherent limitations as mentioned in the Introduction section of this paper. Further these techniques fall flat when the algorithm is analyzed for some unusual data set for which theoretical expectation does not exists. In such a situation the empirical analysis is the only choice.

Let us consider the random variable X which takes the discrete values $x_k = (-1)^k 2^k/k$, (k=1, 2, 3........), with probabilities $p_k = 2^{-k}$. Here we get
$\sum_{k=1}^{\infty}(x_k p_k) = \sum_{k=1}^{\infty}(-1)^k/k = -[1 - 1/2 + 1/3 - 1/4 + .....] = \log_e 2$ and $\sum_{k=1}^{\infty}(|x_k| p_k) = \sum_{k=1}^{\infty} 1/k$ which is a divergent series. Hence in this case expectation does not exist.
Using the inverse cdf technique we have:
$F_X(x) = P(X \leq x) = \sum_{x=x_0}^{r} P(X = x)$, $F_X(x) \sim U[0, 1]$

This unusual data model is simulated over various sample ranges whose regression analysis result is given in table (5) and fig. (6). We have used the quadratic model as a test of linear/$\log_2 n$ goodness of fit. The test is performed by fitting a quadratic model to the data: $y = b_0 + b_1 x + b_2 x_2$, where the regression coefficients $b_0$, $b_1$, and $b_2$ are estimates of the respective parameters. From the regression and ANOVA tables it is evident that the $r^2$ is much higher, the standard error (S) is smaller. The result of coefficient for the various terms are not informative but more importantly, their t values are. The t value for quadratic term is statistically much significant than the other terms, which is an evidence of quadratic nature of the algorithmic behavior for the said data model. This result again refutes the robustness claim of average case behavior of quick sort algorithm.

Table 5. Regression Analysis: t versus n, $n\log_2 n$, $n^2$

```
Regression Analysis: T versus N, NlogN, N^2
The regression equation is
T = 0.010 + 0.000034 N - 0.000002 NlogN + 0.000000 N^2

Predictor       Coef        SE Coef       T      P
Constant      0.0100        0.1024      0.10   0.925
N          0.00003449    0.00003115     1.11   0.311
NlogN     -0.00000212    0.00000190    -1.11   0.308
N^2        0.00000000    0.00000000    34.79   0.000

S = 0.0245745   R-Sq = 100.0%   R-Sq(adj) = 100.0%

Analysis of Variance

Source          DF      SS       MS         F        P
Regression       3    412.47   137.49   227666.93   0.000
Residual Error   6      0.00     0.00
Total            9    412.47

Source    DF    Seq SS
N          1    391.67
NlogN      1     20.07
N^2        1      0.73

Obs      N        T        Fit     SE Fit   Residual   St Resid
 1    20000    0.3057    0.2990   0.0234     0.0066      0.90
 2    40000    0.9044    0.9130   0.0149    -0.0086     -0.44
 3    60000    1.8937    1.9049   0.0149    -0.0112     -0.57
 4    80000    3.2939    3.2858   0.0128     0.0081      0.39
 5   100000    5.0515    5.0611   0.0117    -0.0096     -0.44
 6   120000    7.2734    7.2338   0.0125     0.0396      1.87
 7   140000    9.8092    9.8062   0.0133     0.0030      0.15
 8   160000   12.7451   12.7796   0.0127    -0.0345     -1.64
 9   180000   16.1437   16.1552   0.0131    -0.0115     -0.55
10   200000   19.9518   19.9338   0.0213     0.0180      1.48
```

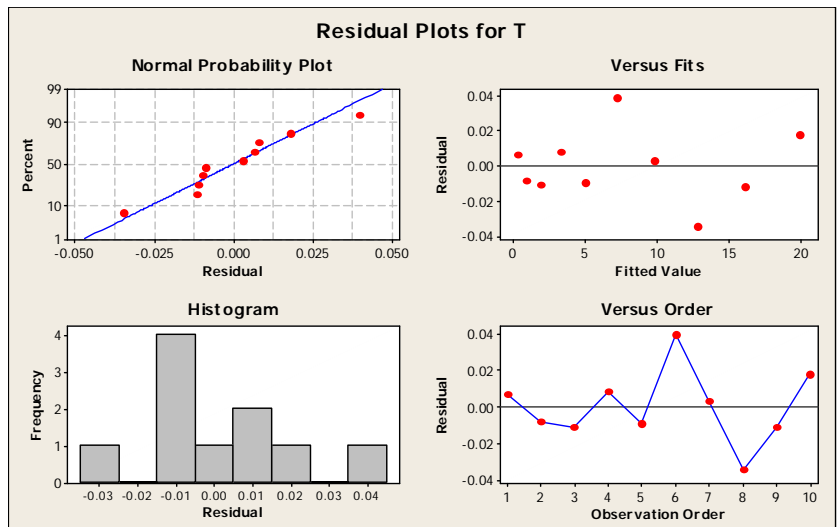
Fig.6. Residual plot for an unusual data model

### 3. CONCLUSIONS

We conclude this paper with the following remarks:

This research paper carefully explores the average case behavior of our candidate algorithm using the unconventional statistical bound and its empirical estimate, the so called 'empirical-O'. The statistical bound has surely much more to offer, as is obvious from our adventure, than its theoretical counterpart. This untraditional and unconventional bound (which actually is an estimate [6]) has the potential to compliment as well as supplement the findings of much practiced mathematical bound.

The statistical analysis performed over empirical data for discrete uniform distribution inputs resulted in several practically interesting patterns. To our surprise we found some complexity data following a very clear linear pattern suggesting an empirical linear model, i.e., $Y_{avg} = O_{emp}(n)$. But interestingly, the proper examination of individual response values put a serious objection on the validity of this proposed complexity model for all practical purposes. These phenomena resulted in conjectures 1 & 2 as given in the main text of this paper. As the last adventure of our tour, we examined the behavior for a non-standard data model for which expectation does not exists theoretically. Based on our statistical analysis result, we have refuted the robustness claim of average case behavior of quick sort algorithm.

The general techniques for simulating the continuous and discrete as well as uniform and non uniform random variables can be found in [19]. For a comprehensive literature on sorting, see references [1] [20]. For sorting with emphasis on the input distribution, [21] may be consulted.

The concept of mixing operations, as is done inherently by experimental approach, of different types is not completely a new idea. In the words of Horowitz et. al. "Given the minimum utility of determining the exact number of additions, subtractions, and so on, that are needed to solve a problem instance with characteristics given by n, we might as well lump all the operations together (provided that the time required by each is relatively independent of the instance characteristics) and obtain a count for the total number of operations" [22]. What is new with our approach is that: instead of count we prefer to work with weights and think of a conceptual bound based on these weights, which relatively is a more scientific approach as it is well known that different operations might take different amount of actual CPU times. The role of weighted count becomes more prominent when the operations in consideration differ drastically with respect to the actual consumed time. The credibility of the statistical bound estimate depends on the design and analysis of our computer experiment in which the response variable is program run time [6].

Our prime objective behind this paper is to convince its reader about the existence of weight based statistical bound. We strongly recommend use of empirical-O for arbitrary algorithms having significant performance gap, as in the present case, between their theoretical average and worst case bounds. The use of empirical-O is also recommended for algorithms in which identifying the key operation itself is a non trivial task. Conceding to the fact that the field of count based theoretical analysis is quite saturated now, hopefully the community of theoretical computer scientists would find our approach a systematic and scientific one, and hence would appreciate our statistical findings.

**APPENDIX**

Definition: Statistical bound (non-probabilistic)

If $w_{ij}$ is the weight of (a computing) operation of type i in the j$^{th}$ repetition (generally time is taken as a weight) and y is a "stochastic realization" (which may not be stochastic [6]) of the deterministic $T = \sum 1 \cdot w_{ij}$ where we count one for each operation repetition irrespective of the type, the statistical bound of the algorithm is the asymptotic least upper bound of y expressed as a function of n where n is the input parameter characterizing the algorithm's input size. If interpreter is used, the measured time will involve both the translation time and the execution time but the translation time being independent of the input will not affect the order of complexity. The deterministic model in that case is $T = \sum 1 \cdot w_{ij} +$ translation time. For parallel computing, summation should be replaced by maximum. **Empirical O** (written as O with a subscript emp) is an empirical estimate of the statistical bound over a finite range, obtained by supplying numerical values to the weights, which emerge from computer experiments [6].

Empirical O can also be used to estimate a mathematical bound when theoretical analysis is tedious, with the acknowledgement that in this case the estimate should be count based and operation specific.